\documentclass[twocolumn,showpacs,preprintnumbers,amsmath,amssymb]{revtex4}

\usepackage{graphicx}
\usepackage{dcolumn}
\usepackage{bm}

\begin{document}

\title{Inter-edge strong-to-weak scattering evolution at a constriction in the fractional quantum Hall regime}
\author{Stefano Roddaro}
\author{Vittorio Pellegrini}
\author{Fabio Beltram}
\affiliation{NEST-INFM, Scuola Normale Superiore, Piazza dei Cavalieri 7, I-56126 Pisa, Italy}
\author{Giorgio Biasiol}
\author{Lucia Sorba}
\affiliation{NEST-INFM and Laboratorio Nazionale TASC-INFM, Area Science Park, I-34012 Trieste, Italy}

\date{\today}

\begin{abstract}
Gate-voltage control of inter-edge tunneling at a split-gate constriction in the fractional quantum Hall regime is reported. Quantitative agreement with the behavior predicted for out-of-equilibrium quasiparticle transport between chiral Luttinger liquids is shown at low temperatures at specific values of the backscattering strength. When the latter is lowered by changing the gate voltage the zero-bias peak of the tunneling conductance evolves into a minimum and a non-linear quasihole-like characteristic emerges. Our analysis emphasizes the role of the local filling factor in the split-gate constriction region.
\end{abstract}

\maketitle
Scattering amplitudes between highly-correlated electronic states are driven by a complex range of fundamental properties and are providing some of the most fascinating manifestations of electron-electron interaction effects in condensed-matter \cite{Chang,Koutouza}. Two-dimensional electron systems (2DES) under the application of magnetic fields (B) and at low temperatures are the ideal experimental arena where these phenomena can be induced and experimentally studied. This extreme quantum limit is characterized by integer and fractional quantum Hall (QH) states \cite{QHE1}. For integer or peculiar fractional ratios of the charge ($n$) and magnetic-flux ($n_{\phi}$) densities (i.e. the filling factor $\nu = n/n_{\phi} = n\cdot h/eB$) the 2DES becomes insulating and charges can only propagate in chiral one-dimensional (1D) states at the edges of the QH liquid. Wen demonstrated that at fractional filling factors these 1D channels lead to a remarkable realization of non-fermionic states \cite{Wen1,Wen2,Wen3} identified as chiral Luttinger liquid (CLL). Several theoretical investigations predicted non-linear tunneling between two such non-Fermi liquids or between a metal and a CLL \cite{Wen1,Wen2,Wen3,KaneFisher,Fendley1,Fendley2, Fendley3,Raimondi,Saleur03,Mandal02}. These results motivate an intense on-going experimental effort in QH systems \cite{Grayson,Yang03,Chang}.  
\par
Experimentally, tunneling between two CLLs (or inter-edge tunneling) can be induced at a quantum point contact (QPC) constriction defined by gating \cite{Milliken, Goldman, Glattli, Heiblum0, Heiblum1, Roddaro1, Roddaro2, Heiblum2}. The split-gate QPC has two main effects. By locally depleting the 2DES it controls the edge separation and consequently the inter-edge interaction strength. Very importantly it also modifies the local filling factor ($\nu^*$). $\nu^*$, in fact, can be different from the bulk filling factor ($\nu$). At gate-voltage ($V_g$) values corresponding to the formation of the constriction (2D-1D threshold) $\nu^*$ is still equal to $\nu$. By further reducing $V_g$ the local filling $\nu^*$ decreases and becomes zero at pinch-off.
\par 
Two separate inter-edge scattering regimes can be identified \cite{KaneFisher}. In the strong backscattering limit the constriction is approaching pinch-off ($\nu^* \approx 0$): scattering is associated to tunneling of electrons between two disconnected QH regions separated by the QPC. For simple fractions such as $\nu = 1/q$ where $q$ is an odd integer, theory predicts that when the tunneling voltage $V_T$ (voltage difference between the two fractional edge states) goes to zero the tunneling current $I_T$ vanishes as $V^{2q-1}_{T}$. In the opposite limit of weak backscattering, the QH fluid is weakly perturbed by the constriction ($\nu^{*}\approx \nu$) and tunneling is associated to scattering of Laughlin quasiparticles with fractional charge $e/q$. In this case and in the T\,=\,0 limit, the tunneling current diverges as $V_T$ tends to zero. At small but finite temperatures and below a critical tunneling voltage $I_T$ reverts to a linear Ohmic behavior. This leads to a zero-bias peak in the differential tunneling conductance $dI_{T}/dV_T$ whose width is proportional to $q\cdot kT/e$. These theoretical suggestions prompted our recent experiments where we observed an unexpected suppression of the tunneling conductance in the low-temperature weak-backscattering limit and the appearance of the inter-edge tunneling zero-bias peak only above a critical value of temperature \cite{Roddaro1, Roddaro2}. This low-temperature suppression has been recently ascribed to inter-edge interactions across the split-gate \cite{Papa04}. 
\par
The crossover between strong and weak regime, and the role of temperature are non-equilibrium quantum transport phenomena largely unexplored experimentally. Fendley et al. \cite{Fendley1} were the first to provide a unified theoretical framework of non-equilibrium transport between CLLs applicable to these different regimes. They demonstrated the existence of an exact duality between weak and strong backscattering i.e. between electron and Laughlin quasiparticle tunneling. \cite{Fendley1,Fendley2,Fendley3}. Recent microscopic calculations emphasized the impact of electron-electron interactions \cite{Mandal02,Mandal01}.  
\par
In this letter we show the experimental evolution of the out-of-equilibrium inter-edge tunneling conductance $dI_T/dV_T$ for different values of $V_g$ and temperature. Particular attention is given to the case $\nu = 1/3$ in the bulk. We find that at $\nu^{*} = 1/5$ the $dI_T/dV_T$ versus $V_T$ characteristic displays a sharp zero-bias peak at low temperatures and two well-resolved minima at positive and negative $V_T$ values. 
Both width and amplitude of the zero-bias peak are found to saturate for temperatures below $T = 100$\,mK.
The shape of the tunneling conductance and its temperature dependence are successfully compared with the predictions by Fendley et al. for non-equilibrium transport through a QPC in a Luttinger liquid \cite{Fendley1,Fendley2}. Our experiments and analysis provide further evidence for the non Fermi-liquid nature of fractional QH edge states. 
\par
As the backscattering strength is lowered by changing $V_g$ we find first a suppression of the non-linearity of the tunneling conductance at $\nu^* \approx  1/4$, and then a quasihole-like characteristic with a zero-bias minimum at $V_g$ corresponding to $\nu^{*} = 2/7$. When $\nu^* = \nu = 1/3$ we recover a zero-bias minimum in agreement with previous results \cite{Roddaro1,Heiblum2}. A similar evolution is also found at a bulk filling factor $\nu = 1$, in this case centered on $\nu^{*} = 1/2$. This latter result unambiguously establishes the role of the local filling factor $\nu^*$ in determining the non-linear inter-edge tunneling behavior, even in a configuration where the bulk is a Fermi-liquid state. We believe that the observed values of $\nu^*$ associated to the peak-to-minimum crossover and yielding a Fendley-like lineshape of the differential tunneling conductance point at an interpretation in term of particle-hole conjugation around the metallic state of composite fermions. Andreev-like processes of fractional quasiparticles at the interface between the bulk and the constriction region \cite{chamon} and intra- and inter-edge interaction effects \cite{Mandal01,Mandal02,Papa04} could also play a significant role.  
\par
\begin{figure}
\includegraphics[width=7.6cm]{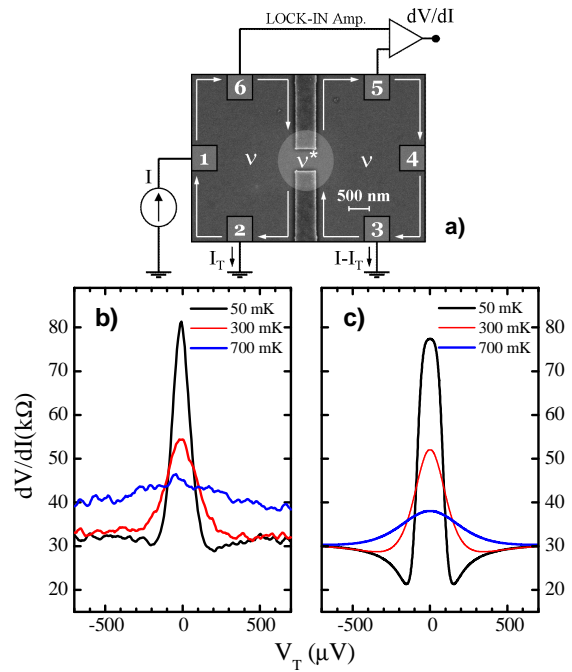}
\caption{\label{fig:0} a) Scanning electron microscope image of the quantum point contact (QPC) and set-up of  differential conductance. The geometrical sizes of the QPC constriction are $600\,{\rm nm}$ wide and $500\,{\rm nm}$ long. Bias current $I$ is injected at the contact $1$ and partially reflected at the constriction. The backscattering current $I_{T}$ leaves the device through the contact $2$, the transmitted one ($I-I_{T}$) through the contact $3$. The measurement is performed in a four-wire scheme. The resistance drop $dV/dI$ is measured between contacts $5$ and $6$. b) Experimental $dV/dI$ tunneling curves as a function of temperatures. $I_{ac} = 20pA$. c) Calculated $dV/dI$ curves as a function of temperature according to Refs.\cite{Fendley1,Fendley2}.}
\end{figure}

 
The measured devices were processed from a $100\,{\rm nm}$ deep GaAs/Al$_{0.1}$Ga$_{0.9}$As heterojunction with carrier density $n = $\,7-9$\times10^{10}\,{\rm cm^{-2}}$ (depending on the cooldown procedure) and mobility always exceeding $10^{6}\,{\rm cm^2/Vs}$. QPC gates were fabricated by e-beam lithography, metallization and lift-off. Figure \ref{fig:0}a shows a scanning electron microscopy image of the QPC superimposed to the multiterminal configuration used. Measurements were carried out by injecting a current $I$ with both ac and dc components into contact $\emph 1$. This current is partially reflected at the constriction: the backscattered fraction ($I_T$) is collected by Ohmic contact $\emph 2$ while the transmitted one ($I-I_{T}$) is collected by contact $\emph 3$. With this configuration, the potential difference between the two edges propagating towards the constriction is given by $V_T = \rho_{xy}I$ 
($\rho _{xy} = h/\nu e^{2}$). Finite-bias phase-locked four-wire measurements were performed with the ac component of the current down to 20\,pA. When the 2DES outside the QPC is in a QH state, the longitudinal-resistance drop across the constriction ($dV/dI$, the quantity measured in the experiment) is related to $dI_T/dV_T$ through the relation: 
\begin{equation} \label{eqS}
\frac{dV}{dI} = \rho_{xy}\frac{dI_{T}}{dI}=\rho_{xy}^2\frac{dI_{T}}{dV_T}. 
\end{equation}
Given the direct proportionality between these two quantities, in what follows the differential tunneling conductance characteristics will be presented as $dV/dI$ resistance curves. 
The mismatch between bulk and constriction filling factors yields an additional $V_T$-independent longitudinal resistance drop due to the extra back-scattered Landauer-Buttiker current. In the set-up adopted in our experiments this mismatch leads to $\frac{dV}{dI}|_{BG}=h/\nu e^2\left[1-\nu^*/\nu\right]$. This relation was used to estimate $\nu^{*}$ from the measured resistance at large $V_T$.
\par
Figure 1b shows representative finite-bias $dV/dI$ measurements at a background value $\frac{dV}{dI}|_{BG} = 32$\,k$\Omega$, i.e. $\nu^{*} = 1/5$. At low temperatures the tunneling conductance displays a sharp (full width at half maximum $\approx$\,200\,$\mu$V) zero-bias peak and two minima at positive and negative voltage bias (the peak resistance value is set by the quantized transverse resistance $3h/e^{2} = 77.4$\,k$\Omega$ \cite{Note02}). The tunneling conductance presents a marked temperature dependence, and the non-linearity  altogether disappears for T exceeding 700-800\,mK. These data can be analyzed within the framework proposed by Fendley et al. \cite{Fendley2}. Figure 1c, in particular, reports a set of calculated differential inter-edge tunneling characteristics at filling factor 1/5 \cite{Fendley2}. To allow the comparison with the experimental data, a constant background of 32 k$\Omega$ was added to the calculated curves. The only free parameter in the calculation is the so-called impurity or point-contact interaction strength $T_B$ and the best agreement with the experimental data was found for $T_B$\,=\,500\,mK. The observed agreement strengthens the interpretation of the non-linear tunneling curve in terms of quasiparticle tunneling between fractional quantum Hall edge states at filling factor 1/5. It is worth noting that the two lateral minima present at the lowest temperature are purely non-equilibrium transport effects in the inter-edge tunneling conductance.
\begin{figure}
\includegraphics[width=7.6cm]{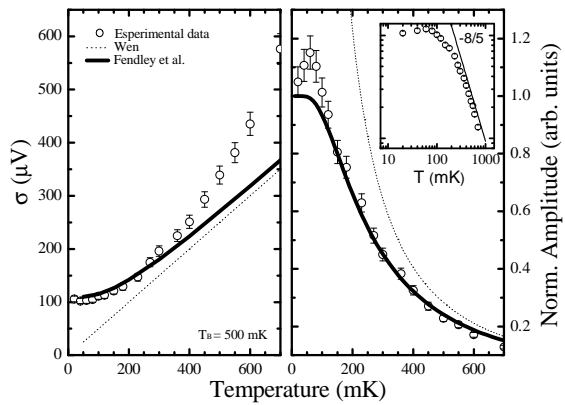}
\caption{\label{Fig03} (a) Standard deviation parameter $\sigma$ of the gaussian fit to the zero-bias tunneling conductance peak verus temperature. Experimental data (open circles), theoretical calculation at filling factor 1/5 following Wen (Ref.\cite{Wen1} dotted line), and Fendley et al. (Ref.\cite{Fendley2} solid line) with $T_B$ = 500mK. (b) Same as in (a) but for the peak intensity (after substraction of the background value of 32k$\Omega$ and normalized to the tranverse resistance value $3h/e^{2} = 77.4k\Omega$). The inset reports the experimental points in a log-log scale together with a straight line with slope -8/5.}
\end{figure}
\par
Further support to this interpretation stems from the temperature dependence of width and intensity of the zero-bias tunneling peak. Experimental data (open circles) are plotted in Fig.\,2 together with the results of the theoretical prediction in the weak-backscattering limit \cite{Wen2,Wen3} (dotted lines) and with the exact results of Fendley et al. \cite{Fendley1,Fendley2} (solid lines). We should like to emphasize the low- and high-T behavior in Fig.\,2. At low-T the weak back-scattering theory predicts a width proportional to $kT/\nu^{*}e$, vanishingly small when T\,$\rightarrow$\,0. On the contrary the exact non-equilibrium results yield a saturation below $T\approx$\,100\,mK in agreement with our experimental results \cite{note03}. A similar saturation was found for the intensity of the zero-bias tunneling peak. This behavior signals the evolution of the tunneling characteristics from the weak- to the strong-backscattering regime as T is lowered. In the high-T weak-backscattering regime, on the other hand, the T-dependence of the peak intensity is compatible with $T^{-8/5}$ (see the inset to Fig.3b where the experimental data are plotted in log-log scale together with a straight line with slope -8/5) consistently with the CLL prediction of $T^{(2\cdot \nu^{*}-2)}$ at $\nu^{*} = 1/5$. 
\par
\begin{figure}
\includegraphics[width=7.6cm]{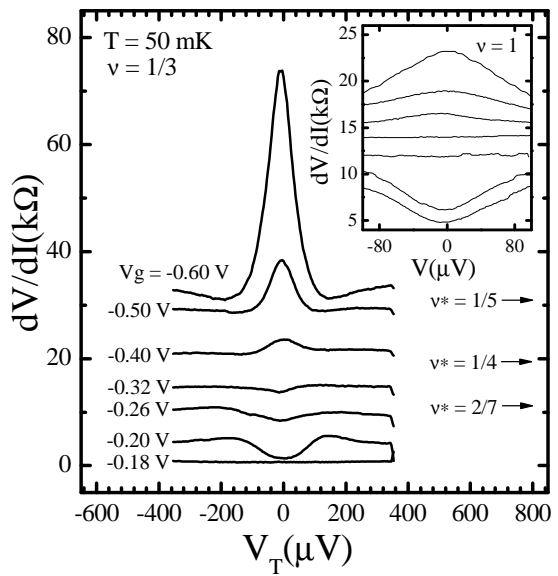}
\caption{\label{fig:02} Evolution of tunneling conductance versus gate volatge at T=50mK and $\nu $ =1/3. The background resistance values associated to relevant filling factors inside the constriction region $\nu^*$ are indicated by the arrows. Inset shows the behavior at the bulk filling factor $\nu $= 1. Here the crossover occurs at $\approx 12$ k$\Omega$ that corresponds to $\nu ^*$ = 1/2 and the saturation is at the transverse resistance $h/e^{2} = 25.8k\Omega$. The excitation current is $I_{ac} = 200pA$.}
\end{figure}

Let us move on to the evolution of inter-edge tunneling conductance as a function of $V_g$. Figure~\ref{fig:02} reports representative results obtained with a relatively-high excitation current $I_{ac} = 200$\,pA. Two qualitatively different behaviors emerge: for high backscattering strength (high $|V_g|$ values) conductance curves display a maximum at zero bias (see also Fig. 1b), for lower $|V_g|$ values the maximum evolves into a minimum. This behavior was consistently observed at different charge densities (and magnetic fields). In all measurements we found that the minimum-to-maximum transition occurs at a resistance value of about $20\,{\rm k\Omega}$ where a flat linear characteristic is found. It is intriguing to note that this resistance value corresponds to $\nu^{*} = 1/4$. Here a Fermi-liquid state of composite fermions with four flux quanta $h/e$ attached is realized within the constriction region.  A similar evolution is observed at $\nu$ = 1 (see inset of Fig.\,3): the crossover in this case occurs at $\nu^{*}$ =1/2. This latter result highlights the impact of the local filling factor $\nu^*$ on the tunneling characteristics. 

A careful analysis of the tunneling conductance characteristics under low excitation current (20\,pA) in the gate-voltage region of the minima reveals an unexpected symmetric behavior. While, in fact, the lineshapes of maxima and minima change significantly as the gate-voltage is varied in a manner that can not be modeled in a simple way, at the specific background resistance value of about 11\,k$\Omega $ ($\nu ^{*} = 2/7$) we observed a sharp zero-bias dip with  lineshape similar to the one observed at $\nu^* = 1/5$. Figure 4 compares the evolution of the tunneling conductance peak ($\nu^{*} = 1/5$) and dip ($\nu^{*} = 2/7$) as a function of temperature. The behavior in the high temperature limit is associated to the breakdown of the QH state outside the QPC: in both cases the contribution of this additional backscattering simply increases the overall resistance drop. As expected the weaker QH $2/7$ state reflects into a more pronounced temperature dependence. 

\begin{figure}
\includegraphics[width=7.6cm]{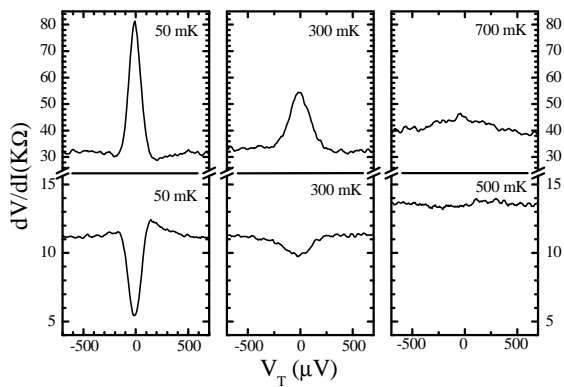}
\caption{\label{fig:01} Top panels: Evolution of the differential tunneling conductance characteristics corresponding to $\nu^*$ = 1/5 (background resistance value of $\approx 32$ k$\Omega$) at three different values of the temperature. Bottom panels: same as in upper panels but for the background resistance of 11k$\Omega $ ( corresponding to $\nu^*$ = 2/7). The excitation current is $I_{ac} = 20pA$.}
\end{figure}

The symmetric evolution of the two lineshapes is truly intriguing. Given the values of $\nu ^*$'s it is tempting to link these data to particle-hole conjugation around the metallic state of composite fermions at 1/4. In this framework the tunneling peak could be related to quasi-particle tunneling between $\nu^* = 1/5$ edge states and the dip would be due to quasi-hole tunneling at $\nu^*=2/7$, the latter leading to an increase of the total transmission coefficient at the QPC constriction (i.e. reduction of the measured resistance drop $dV/dI$). In the Fermi-liquid state corresponding to $\nu^{*} = 1/4$ the $V_T$-dependent non-linear tunneling current vanishes as observed experimentally. The complex structure of edge states in the rather smooth potential profile of the split-gate QPC should be taken into account \cite{halperin}. Moreover quasi-particle Andreev-like processes due to the mismatched filling factors at the QPC \cite{chamon} and interaction effects intra-edge and across the split-gate \cite{Mandal01,Papa04} may also play a role. Further experimental and theoretical analysis is therefore needed. 
\par
In conclusion we reported the evolution of inter-edge scattering at a split-gate QPC constriction when the bulk is at $\nu=1/3$. Gate bias allows to control the inter-edge coupling by changing both the inter-edge distance and the filling factor $\nu ^*$ within the QPC region. At a local filling factor $\nu^{*} = 1/5$ and $T = 50$\,mK we observed a zero-bias differential tunneling-conductance peak consistent with the prediction of out-of-equilibrium quasi-particle transport between CLLs. The gate-voltage-induced evolution of tunneling around $\nu^{*} = 1/4$ suggests an unexpectedly complex phenomenology in quasi-particle transport through a QPC in the QH regime.      
\par
We are grateful to M. Grayson, J.K. Jain, A.H. MacDonald, E. Papa, R. Raimondi, B. Trauzettel, G. Vignale for discussions and suggestions. This work was supported in part by the Italian Ministry of University and Research under FIRB RBNE01FSWY and by the European Community's Human Potential Programme under contract HPRN-CT-2002-00291 (COLLECT).


\begin{thebibliography}{9}
\bibitem{note03}
{The current modulation in our experiments yields a threshold temperature of $V_{T,ac}\approx \rho_{xy}I_{ac}\approx 10 mK$ well below our lowest accessible temperature.}
\end{thebibliography}
\end{document}